\documentclass[twocolumn,showpacs,aps]{revtex4}
\usepackage{epsfig}
\usepackage{amsbsy}
\usepackage{amsmath}
\usepackage{graphicx}
\usepackage{latexsym}

\tighten

\begin{document}

\title{Spin swapping operator as an entanglement witness for quantum
Heisenberg spin-$s$ systems}
\author{Guang-Ming Zhang$^{1,3}$ and Xiaoguang Wang$^{2,3}$}
\affiliation{$^{1}$Department of Physics, Tsinghua University, Beijing 100084, China;\\
$^{2}$Zhejiang Institute of Modern Physics, Department of Physics, Zhejiang
University, Hanzhou 310027, China.\\
$^{3}$Department of Physics and Center of Theoretical and Computational
Physics, The University of Hong Kong, Pokfulam Road, Hong Kong, China.}
\date{\today}

\begin{abstract}
Using the SU($N$) representation of the group theory, we derive the general
form of the spin swapping operator for the quantum Heisenberg spin-$s$
systems. We further prove that such a spin swapping operator is equal to the
spin singlet pairing operator under the partial transposition. For SU(2)
invariant states, it is shown that the expectation value of the spin
swapping operator and its generalizations, the permutations, can be used as
an entanglement witness, especially, for the formulation of observable
conditions of entanglement.
\end{abstract}

\maketitle

\section{Introduction}

Entanglement is one of the most intriguing properties of quantum physics and
the key ingredient of quantum information and processing. To determine the
existence of entanglement, partial transposition of the density matrix is
introduced\cite{peres,horodecki1996}. In $2\times 2$ and $2\times 3$
dimensional Hilbert spaces, the requirement of positive partial
transposition (PPT) represents a strong necessary and sufficient criterion
for the separability of states, the so-called Peres-Horodecki criterion\cite%
{peres,horodecki1996}. A useful entanglement measure for higher dimensions,
the negativity, is defined by the sum of absolute value of negative
eigenvalues of the partial transposed density matrix\cite{Neg} though such a
criterion of entanglement is no longer sufficient.

Recently it has been realized that symmetries in the mixed states play an
important role in characterizing the entanglement properties \cite%
{werner1989,horodecki1999,werner2001,schliemann,breuer}. For the SU(2)
invariant states in dimensions $2\times L$, $3\times M$, and $4\times 4$,
respectively, the Peres-Horodecki criterion has been proved to be necessary
and sufficient\cite{schliemann,breuer,breuer1}, where $L=2j+1$ with
arbitrary spin-$j$ and $M=2j^{\prime }+1$ with $j^{\prime }$ being integer.

To analyze the general structure of the state space for bipartite $N\times N$
quantum systems, we can regard the subsystems as quantum Heisenberg spin-$s$
systems ($N=2s+1$) and transform according to an SU($N$) irreducible
representation of the group theory. By the requirement of SU(2) invariance,
we can substantially reduce the dimensionality of the state space, and the
entanglement criteria become easy to be handled analytically.

On the other hand, the entanglement properties in Heisenberg spin systems
have received much attention\cite{M_Nielsen}-\cite{QPT_GVidal}. For the
quantum spin-1/2 system, there is an SU(2) invariant operator, i.e., the
swapping operator
\begin{equation}
\mathbf{S}_{i,j}=2\mathbf{s}_{i}\mathbf{\cdot s}_{j}+\frac{1}{2},
\end{equation}%
which switches the spin states on the sites of $i$ and $j$. Such a swapping
operator satisfies $\mathbf{S}_{i,j}^{2}=1$ and $\mathbf{S}_{i,j}^{\dagger }=%
\mathbf{S}_{i,j}$. Therefore, every SU(2) invariant density matrix can be
expressed as $\rho _{i,j}=b+c\mathbf{S}_{i,j}$ with suitable real parameters
$b$ and $c$. Actually, one can simply use a single parameter $\langle
\mathbf{S}_{i,j}\rangle =\mathrm{Tr}(\rho _{i,j}\mathbf{S}_{i,j})$, which
ranges from $-1$ to $1$, to describe these SU(2) invariant states. It is
important to notice that for an SU(2) invariant state, the condition $%
\left\langle \mathbf{S}_{i,j}\right\rangle <0$ has been proved to be \textit{%
sufficient }and\textit{\ necessary} for entanglement\cite{xgwang}. There
also exists a simple relation between the concurrence\cite{Conc},
quantifying two-qubit entanglement, and the expectation value of the
swapping operator with respect to the density matrix $\rho _{i,j}$
\begin{equation}
\mathcal{C}_{ij}=\max (0,-\left\langle \mathbf{S}_{i,j}\right\rangle ).
\end{equation}%
However, for $s>1/2$, the operator $2\mathbf{s}_{i}\mathbf{\cdot s}_{j}+%
\frac{1}{2}$ can no longer be regarded as a spin swapping operator, because
the SU(2) description is not the faithful fundamental representation for the
quantum spin-$s$ operators.

In this paper, based on the SU($N$) representation of the group theory, we
will first derive the general form of the spin swapping operator for the
quantum Heisenberg spin-$s$ system. Then it will be proved that the partial
transposed swapping operator is just equal to the singlet pairing operator
defined in the tensor product space of the fundamental SU($N$)
representation and its conjugate one. For an SU(2) invariant spin-$s$
system, we will show that the expectation value of the swapping operator
gives rise to the leading contribution to the negativity expressed in terms
of the Wigner 6-j symbol. Generalized to the many-body particle states, it
will be concluded that the expectation values of the swapping and its
generalizations, the permutations, can be used as an entanglement
witnesses~(EWs) \cite{EW_Horodecki,EW_Terhal,EW_Lewenstein}, and are useful
for the formulation of observable conditions of entanglement.

\section{Swapping and singlet projector for quantum Heisenberg spin-$s$
systems}

\subsection{Spin swapping operator}

To describe a spin-$s$ operator quantum mechanically, we use the good
quantum numbers: $\mathbf{s}^{2}=s(s+1)$ and $s_{z}=-s$, $-s+1$, $...$, $s$.
The dimensionality of the local Hilbert space is thus $N=2s+1$. It is
natural to introduce an SU($N$) fundamental symmetry group with generators
in terms of bosons/fermions\cite{auerbach}
\begin{equation}
F_{\mu }^{\nu }(i)=a_{i,\mu }^{\dagger }a_{i,\nu },
\end{equation}%
where $\mu $ and $\nu $ denote the spin projection indices from $1,2,...,2s+1
$, and $i$ denotes the site. By using the commutation/anticommutation
relations,%
\begin{eqnarray}
\left[ a_{i,\mu },a_{j,\nu }\right] _{\mp } &=&\left[ a_{i,\mu }^{\dagger
},a_{j,\nu }^{\dagger }\right] _{\mp }=0,  \notag \\
\left[ a_{i,\mu },a_{j,\nu }^{\dagger }\right] _{\mp } &=&\delta
_{i,j}\delta _{\mu ,\nu },
\end{eqnarray}%
we can prove that the generators satisfy the following commutation relation
of the SU($N$) Lie algebra\cite{group-theory}%
\begin{equation}
\left[ F_{\mu }^{\nu }(i),F_{\mu ^{\prime }}^{\nu ^{\prime }}(j)\right]
=\delta _{i,j}\left( \delta _{\nu ,\mu ^{\prime }}F_{\mu }^{\nu ^{\prime
}}(i)-\delta _{\mu ,\nu ^{\prime }}F_{\mu ^{\prime }}^{\nu }(i)\right) .
\end{equation}%
Accordingly, the corresponding spin operator is expressed as
\begin{equation}
s_{i}^{\alpha }=\sum_{\mu ,\nu }a_{i,\mu }^{\dagger }T_{\mu \nu }^{\alpha
}a_{i,\nu },
\end{equation}%
where $T^{\alpha }$ ($\alpha =x,y,z$) are the corresponding $N\times N$
matrices for the quantum spin-$s$ operator. We can also prove that the
commutation relations of the SU(2) Lie algebra are also satisfied when
inserting the expressions of the spin-$s$ operators. In order to fix the
magnitude of the quantum spins $\mathbf{s}_{i}^{2}=s(s+1)$, a local
constraint $\sum_{\mu }a_{i,\mu }^{\dagger }a_{i,\mu }=1$ has to be imposed
as well.

With the help of these SU($N$) generators, the general swapping operator
between any two sites with $N$ local states each can be constructed as%
\begin{equation}
\mathbf{S}_{i,j}=\sum_{\mu ,\nu }F_{\mu }^{\nu }(i)F_{\nu }^{\mu
}(j)=\sum_{\mu ,\nu }a_{i,\mu }^{\dagger }a_{j,\nu }^{\dagger }a_{j,\mu
}a_{i,\nu },
\end{equation}%
which is the \textit{unique} invariant operator under the local SU($N$)
unitary transformation. In analogy to the Werner states\cite{werner1989}, we
can define the SU($N$)$\times $SU($N$) invariant states as follows%
\begin{eqnarray}
\rho _{i,j} &=&p\rho _{-}+(1-p)\rho _{+},  \notag \\
\rho _{\pm } &=&\frac{1}{N(N\pm 1)}(1\pm \mathbf{S}_{i,j}),
\end{eqnarray}%
where $p=(1+\langle \mathbf{S}_{i,j}\rangle )/2$ is positive parameter
ranging from $0$ to $1$. Actually, the expectation value of this generalized
swapping operator $\langle \mathbf{S}_{i,j}\rangle =$Tr$(\rho _{i,j}\mathbf{S%
}_{i,j})$, which still ranges from $-1$ to $1$, can be used to describe
these SU($N$)$\times $SU($N$) invariant states. We will further prove that
the condition $\langle \mathbf{S}_{i,j}\rangle <0$ is sufficient for
entanglement.

In order to make the swapping operator as a useful EW, it is essential to
rewrite this SU($N$) swapping operator in terms of cumulants of the original
SU(2) spin-$s$ operators. We first notice that
\begin{equation}
\mathbf{s}_{i}\mathbf{\cdot s}_{j}=\frac{1}{2}\left[ (\mathbf{s}_{i}+\mathbf{%
s}_{j})^{2}-2s(s+1)\right] .
\end{equation}%
The Hilbert space is thus given by the tensor product space of two quantum
spins, and can be decomposed into a sum of irreducible representations in
terms of projection operators
\begin{equation}
\mathbf{P}_{F}=\sum_{M=-F}^{F}|F,M\rangle \langle F,M|,
\end{equation}%
where $F=0,1,2,...,2s$ denotes the total spin quantum number, $\mathbf{P}%
_{F} $ is the projection operator of the total spin-$F$ channel, and $%
|F,M\rangle $ corresponds to the irreducible subspace of the tensor product
representation for a fixed $F$. Therefore, a set of relations can be derived
as%
\begin{eqnarray}
(\mathbf{s}_{i}\mathbf{\cdot s}_{j})^{n} &=&\sum_{F=0}^{2s}\lambda _{F}^{n}%
\mathbf{P}_{F},  \notag \\
\lambda _{F} &=&\frac{1}{2}\left[ F(F+1)-2s(s+1)\right] ,
\end{eqnarray}%
where the integer $n=0,1,2,...,2s$. Namely, we have a set of equations for
the projection operators
\begin{eqnarray}
\mathbf{P}_{0}+\mathbf{P}_{1}+\mathbf{P}_{2}+...+\mathbf{P}_{2s} &=&1,
\notag \\
\lambda _{0}\mathbf{P}_{0}+\lambda _{1}\mathbf{P}_{1}+\lambda _{2}\mathbf{P}%
_{2}+...+\lambda _{2s}\mathbf{P}_{2s} &=&\mathbf{s}_{i}\mathbf{\cdot s}_{j},
\notag \\
\lambda _{0}^{2}\mathbf{P}_{0}+\lambda _{1}^{2}\mathbf{P}_{1}+\lambda
_{2}^{2}\mathbf{P}_{2}+...+\lambda _{2s}^{2}\mathbf{P}_{2s} &=&\left(
\mathbf{s}_{i}\mathbf{\cdot s}_{j}\right) ^{2},  \notag \\
&&......  \notag \\
\lambda _{0}^{2s}\mathbf{P}_{0}+\lambda _{1}^{2s}\mathbf{P}_{1}+\lambda
_{2}^{2s}\mathbf{P}_{2}+...+\lambda _{2s}^{2s}\mathbf{P}_{2s} &=&\left(
\mathbf{s}_{i}\mathbf{\cdot s}_{j}\right) ^{2s}.
\end{eqnarray}%
Note that the coefficients in front of the projection operators are of the
form $\lambda _{F}^{n}$, i.e., the corresponding matrix is of the
Vandermonde type with the determinant%
\begin{equation}
\left|
\begin{array}{ccccc}
1 & 1 & 1 & ... & 1 \\
\lambda _{0} & \lambda _{1} & \lambda _{2} & ... & \lambda _{2s} \\
\lambda _{0}^{2} & \lambda _{1}^{2} & \lambda _{2}^{2} & ... & \lambda
_{2s}^{2} \\
.. & .. & .. & ... & .. \\
\lambda _{0}^{2s} & \lambda _{1}^{2s} & \lambda _{2}^{2s} & ... & \lambda
_{2s}^{2s}%
\end{array}%
\right| =\prod\limits_{k<l}(\lambda _{k}-\lambda _{l}).
\end{equation}%
By using the property of the Vandermonde determinant, we can obtain the
general expression for the projection operators in terms of the SU(2) spin-$%
s $ operators
\begin{equation}
\mathbf{P}_{F}=\prod\limits_{\substack{ k=0  \\ \neq F}}^{2s}\left[ \frac{%
\mathbf{s}_{i}\mathbf{\cdot s}_{j}-\lambda _{k}}{\lambda _{F}-\lambda _{k}}%
\right] .
\end{equation}%
Moreover, the general SU($N$) invariant swapping operator can thus be
expressed as%
\begin{equation}
\mathbf{S}_{i,j}=(-1)^{2s}\sum_{F=0}^{2s}(-1)^{F}\mathbf{P}_{F}=\prod\limits
_{\substack{ k=0  \\ \neq F}}^{2s}\left( \frac{\mathbf{s}_{i}\mathbf{\cdot s}%
_{j}-\lambda _{k}}{\lambda _{F}-\lambda _{k}}\right) .
\end{equation}%
Namely, the general spin swapping operator is written as a linear
combination of all projection operators for the spin-$F$ channels with
alternating sign, and $\mathbf{S}_{i,j}$ is symmetric for integer spins and
antisymmetric for the odd-half integer spins when interchanging the spin
states on the sites of $i$ and $j$. Similar expressions for the projections
had appeared in the literature\cite{barber,batchelor-yung}.

As examples, the first four expressions of the general swapping operators
are explicitly written as

i). For $s=1/2$, the above expression gives rise to%
\begin{equation}
\mathbf{S}_{i,j}=2\mathbf{s}_{i}\mathbf{\cdot s}_{j}+\frac{1}{2},
\end{equation}%
which is invariant under the SU(2) unitary transformation.

ii). For $s=1$, the swapping operator is
\begin{equation}
\mathbf{S}_{i,j}=\left( \mathbf{s}_{i}\mathbf{\cdot s}_{j}\right)
^{2}+\left( \mathbf{s}_{i}\mathbf{\cdot s}_{j}\right) -1,
\end{equation}%
which is invariant under the SU(3) unitary transformation.

iii). For $s=3/2$, the swapping operator takes the form%
\begin{eqnarray}
\mathbf{S}_{i,j} &=&\frac{2}{9}\left( \mathbf{s}_{i}\mathbf{\cdot s}%
_{j}\right) ^{3}+\frac{11}{18}\left( \mathbf{s}_{i}\mathbf{\cdot s}%
_{j}\right) ^{2}  \notag \\
&&-\frac{9}{8}\left( \mathbf{s}_{i}\mathbf{\cdot s}_{j}\right) -\frac{67}{32}%
,
\end{eqnarray}%
which is invariant under the SU(4) unitary transformation.

iv). For $s=2$, the swapping operator is expressed as%
\begin{eqnarray}
\mathbf{S}_{i,j} &=&\frac{1}{36}\left( \mathbf{s}_{i}\mathbf{\cdot s}%
_{j}\right) ^{4}+\frac{1}{6}\left( \mathbf{s}_{i}\mathbf{\cdot s}_{j}\right)
^{3}  \notag \\
&&-\frac{13}{36}\left( \mathbf{s}_{i}\mathbf{\cdot s}_{j}\right) ^{2}-\frac{5%
}{2}\left( \mathbf{s}_{i}\mathbf{\cdot s}_{j}\right) -1.
\end{eqnarray}%
which is invariant under the SU(5) transformation.

Thus, the expectation value of the swapping operator $\left\langle \mathbf{S}%
_{i,j}\right\rangle $ can be written in terms of the cumulants of the
quantum spin-$s$ correlators. In solid state physics, the swapping operator
is used to represent the generalized SU($N$) invariant quantum Heisenberg
spin-$s$ model, i.e., $H=J\sum_{\langle i,j\rangle }\mathbf{S}_{i,j}$, to
describe the possible nearest neighbor couplings of magnetic spin-$s$
moments. In one dimension, there exists so-called Bethe ansatz exact solution%
\cite{uimin,lai}. For the antiferromagnetic coupling ($J>0$), the ground
state is a singlet with spin \textit{gapless} excitations\cite{sutherland}.

\subsection{Spin singlet projector}

Among all the projection operators, the singlet projector represents a
maximally entangled state, and its expectation value in some cases has been
used for formulation of necessary and sufficient conditions of entanglement.
In terms of original SU(2) spin-$s$ operators, we have%
\begin{equation}
\mathbf{P}_{ij}=\mathbf{P}_{F=0}=\prod\limits_{k=1}^{2s}\left[ 1-2\frac{%
\mathbf{s}_{i}\mathbf{\cdot s}_{j}+s(s+1)}{k(k+1)}\right] .
\end{equation}%
The corresponding spin singlet state can be projected onto the angular
momentum singlet state%
\begin{equation}
|0,0\rangle =\frac{1}{\sqrt{2s+1}}\sum_{m=-s}^{s}(-1)^{s-m}|s,m\rangle
_{i}\otimes |s,-m\rangle _{j}.
\end{equation}

In particular, the first four expressions for the singlet projectors can be
explicitly written as

i). For $s=1/2$, the singlet operator is
\begin{equation}
\mathbf{P}_{ij}=\frac{1}{4}-\mathbf{s}_{i}\mathbf{\cdot s}_{j}.
\end{equation}%
Then, the swapping operator $\mathbf{S}_{i,j}$ and the singlet projection
operator $\mathbf{P}_{i,j}$ are not independent. They have the following
relation $\mathbf{S}_{i,j}=(1-2\mathbf{P}_{i,j})$. The entanglement
criterion for the SU(2) invariant states $\langle \mathbf{S}_{i,j}\rangle <0$
implies that $\langle \mathbf{P}_{i,j}\rangle >1/2$.

ii). For $s=1$, the singlet projection is given by
\begin{equation}
\mathbf{P}_{i,j}=\frac{1}{3}\left[ \left( \mathbf{s}_{i}\mathbf{\cdot s}%
_{j}\right) ^{2}-1\right] .
\end{equation}

iii). For $s=3/2$, the singlet projection is written as%
\begin{eqnarray}
\mathbf{P}_{i,j} &=&\frac{33}{128}+\frac{31}{96}\mathbf{s}_{i}\mathbf{\cdot s%
}_{j}  \notag \\
&&-\frac{5}{72}(\mathbf{s}_{i}\mathbf{\cdot s}_{j})^{2}-\frac{1}{18}(\mathbf{%
s}_{i}\mathbf{\cdot s}_{j})^{3}.
\end{eqnarray}

iv). For $s=2$, the singlet projection is expressed as%
\begin{eqnarray}
\mathbf{P}_{i,j} &=&-\frac{1}{3}\mathbf{s}_{i}\mathbf{\cdot s}_{j}-\frac{17}{%
180}(\mathbf{s}_{i}\mathbf{\cdot s}_{j})^{2}  \notag \\
&&+\frac{1}{45}(\mathbf{s}_{i}\mathbf{\cdot s}_{j})^{3}+\frac{1}{180}(%
\mathbf{s}_{i}\mathbf{\cdot s}_{j})^{4}.
\end{eqnarray}%
All these singlet projectors display \textit{uniform} SU(2) invariance
superficially, but it will be further proved that a non-uniform higher
symmetry is associated with each singlet projector.

Therefore, the expectation value of the singlet projectors can be also
expressed in terms of the cumulants of the quantum spin-s correlators. In
solid state physics, the singlet pairing projection is also used to
represent another type of the generalized quantum Heisenberg spin-$s$ model,
i.e., $H=-J\sum_{\langle i,j\rangle }\mathbf{P}_{i,j}$, to describe the
nearest neighbor couplings of the magnetic spin-$s$ moments. In one
dimension, an exact solution has been found based on Temperley-Lieb algebra%
\cite{barber}. Moreover, in the case of $J>0$, the ground state is a
dimerized-like singlet state with \textit{gapful} spin excitations\cite%
{batchelor-yung}.

\subsection{Relation between swapping and singlet pairing operators}

According to group theory\cite{group-theory}, for an SU($N$) Lie group with $%
s>1/2$, two kinds of spinors (upper and lower) can actually be defined. The
lower spinor transforms according to the SU($N$) fundamental representation,
while the upper spinor transforms according to the SU($N$) conjugate
representation. More importantly, the conjugate representation is in general
independent of the fundamental representation. Only for $s=1/2$ ($N=2$), due
to the presence of an additional \textit{particle-hole} symmetry, these two
representations are equivalent to each other\cite{group-theory}.

The generators of the SU($N$) conjugate representation is defined by\cite%
{auerbach}%
\begin{equation}
\widetilde{F}_{\mu }^{\nu }(i)=a_{i,\nu }^{\dagger }a_{i,\mu },
\end{equation}%
where $\mu $ and $\nu $ denote the spin projection indices from $1,2,...,2s+1
$, and $i$ denotes the site. By using the commutation/anticommutation
relations for bosons/fermions, we can prove the following commutation
relation%
\begin{equation}
\left[ \widetilde{F}_{\mu }^{\nu }(i),\widetilde{F}_{\mu ^{\prime }}^{\nu
^{\prime }}(j)\right] =\delta _{i,j}\left( \delta _{\nu ,\mu ^{\prime }}%
\widetilde{F}_{\mu }^{\nu ^{\prime }}(i)-\delta _{\mu ,\nu ^{\prime }}%
\widetilde{F}_{\mu ^{\prime }}^{\nu }(i)\right) ,
\end{equation}%
which also forms an SU($N$) Lie algebra. Consider two quantum spins, i.e.,
the bipartite system. With the help of generators of the SU($N$) fundamental
and its conjugate representations, a singlet pairing operator between two
sites $i$ and $j$ can be constructed as
\begin{equation}
\mathbf{P}_{i,j}^{\prime }=\sum_{\mu ,\nu }F_{\mu }^{\nu }(i)\widetilde{F}%
_{\nu }^{\mu }(j)=\sum_{\mu ,\nu }a_{i,\mu }^{\dagger }a_{j,\mu }^{\dagger
}a_{j,\nu }a_{i,\nu },
\end{equation}%
which is the unique SU($N$)$\times \widetilde{\text{SU(}N\text{)}}$
invariant operator and is positive with norm $d=2s+1$. The corresponding
maximally entangled state is expressed as
\begin{equation}
|0,0\rangle ^{\prime }=\frac{1}{\sqrt{2s+1}}\sum_{m=-s}^{s}|s,m\rangle
_{i}\otimes |s,m\rangle _{j}.
\end{equation}%
In analogy to the so-called symmetric/isotropic states \cite{horodecki1999},
we can define the SU($N$)$\times \widetilde{\text{SU(}N\text{)}}$ invariant
states, and every SU($N$)$\times \widetilde{\text{SU(}N\text{)}}$ invariant
state can be expressed as $\rho _{i,j}=b^{\prime }+c^{\prime }\mathbf{P}%
_{i,j}^{\prime }$ with suitable real parameters $b^{\prime }$ and $c^{\prime
}$, or in terms of a convex combination of two minimal projections%
\begin{equation}
\rho _{1}=\frac{1}{2s+1}\mathbf{P}_{i,j}^{\prime },\rho _{2}=\frac{1}{4s(s+1)%
}(1-\rho _{1}).
\end{equation}

Now we are in the position to establish the relation between the general
spin swapping and the singlet pairing operators. In studying entanglement a
powerful tool, the operation of partial transposition, has been introduced%
\cite{peres,horodecki1996}. The partial transposition of an operator in the $%
N\times N$ product space of a bipartite system is defined in a product basis
by transposing only the indices belonging to the second basis and keeping
those pertaining to the first basis. When applying such a partial
transposition operation to the SU($N$)$\times \widetilde{\text{SU(}N\text{)}}
$ invariant singlet pairing operator, we find a very important relation%
\begin{eqnarray}
\mathbf{P}_{i,j}^{\prime } &=&\sum_{\mu ,\nu }a_{i,\mu }^{\dagger }a_{j,\mu
}^{\dagger }a_{j,\nu }a_{i,\nu },  \notag \\
&\Leftrightarrow &\sum_{\mu ,\nu }a_{i,\mu }^{\dagger }a_{j,\nu }^{\dagger
}a_{j,\mu }a_{i,\nu }=\mathbf{S}_{i,j}.
\end{eqnarray}%
Namely, the partial transpose of the SU($N$)$\times \widetilde{\text{SU(}N%
\text{)}}$ invariant singlet pairing operator is \textit{exactly} equivalent
to the uniform SU($N$)$\times $SU($N$) invariant swapping operator. The
inverse statement also holds true. This is one of the main results of our
present paper. Actually a similar relation exists between the Werner states
and symmetric/isotropic states\cite{werner2001}.

Moreover, Breuer has convincingly demonstrated that \cite{breuer} the
partial transposition is \textit{equivalent} to the partial time reversal
transformation of the quantum Heisenberg spin-$s$ operator. Under such a
partial time reversal transformation, the corresponding SU($N$)$\times
\widetilde{\text{SU(}N\text{)}}$ invariant singlet pairing state \textit{%
exactly} transforms into the singlet state in the fundamental SU($N$)
representation%
\begin{eqnarray*}
|0,0\rangle ^{\prime } &=&\frac{1}{\sqrt{2s+1}}\sum_{m=-s}^{s}|s,m\rangle
_{i}\otimes |s,m\rangle _{j} \\
&\Leftrightarrow &\frac{1}{\sqrt{2s+1}}\sum_{m=-s}^{s}(-1)^{s-m}|s,m\rangle
_{i}\otimes |s,-m\rangle _{j},
\end{eqnarray*}%
implying that the spin singlet projection state defined in the SU($N$)
fundamental representation equal to the spin singlet pairing state defined
by the product of the SU($N$) fundamental and its conjugate representations.
Moreover, the singlet projection operator $\mathbf{P}_{i,j}$ shares the same
symmetry of SU($N$)$\times \widetilde{\text{SU(}N\text{)}}$ displayed by the
singlet paring operator.

\section{Swapping operator and its generalizations as entanglement witnesses}

Formulation of different criteria, which allows one to distinguish in
experiment entangled and disentangled states, is one of the most important
issues in the field of foundations of quantum physics and quantum
information processing. The corresponding studies lead to quick development
of the theory of EWs~\cite{EW_Dowling}-\cite{EW_Stobioska}. An entanglement
witness~\cite{EW_Horodecki,EW_Terhal,EW_Lewenstein} is a Hermitian operator
with a key property that its expectation value on a separable state is
always larger or equal to zero. So, if the expectation value on a state is
less than zero, then the corresponding state is entangled.

\subsection{Swapping and negativity}

Consider a many-body state, we first study the two-spin state, and the
generalization to a many particle spin state is straightforward. Swapping
operator exhibits a uniform SU($N$) symmetry, and we may exploit it to
detect entanglement in a quantum Heisenberg spin-$s$ system. The action of
the swapping on a product state is given by
\begin{equation}
\mathbf{S}_{ij}|\phi _{i}\rangle \otimes |\phi _{j}\rangle =|\phi
_{j}\rangle \otimes |\phi _{i}\rangle .
\end{equation}%
A separable (non-entangled) two-particle reduced density matrix $\rho _{ij}$
is introduced as
\begin{equation}
\rho _{ij}=\sum_{k}p_{k}|\phi _{i}^{k}\rangle \langle \phi _{i}^{k}|\otimes
|\phi _{j}^{k}\rangle \langle \phi _{j}^{k}|,
\end{equation}%
where the coefficients $p_{k}$ are positive real numbers, satisfying $%
\sum_{k}p_{k}=1$, and $|\phi _{i}^{k}\rangle $ is the state for the $i$-th
particle. Evaluating the expectation value of $\mathbf{S}_{i,j}$ on this
separable state, we find that
\begin{align}
\left\langle \mathbf{S}_{i,j}\right\rangle =& \text{Tr}(\mathbf{S}_{i,j}\rho
_{ij})  \notag \\
=& \text{Tr}\left( \sum_{k}p_{k}|\phi _{j}^{k}\rangle \langle \phi
_{i}^{k}|\otimes |\phi _{i}^{k}\rangle \langle \phi _{j}^{k}|\right)  \notag
\\
=& \sum_{k}p_{k}|\langle \phi _{i}^{k}|\phi _{j}^{k}\rangle |^{2}\geq 0.
\end{align}%
This inequality is fulfilled for all separable states, and it directly
follows that any state with $\left\langle \mathbf{S}_{i,j}\right\rangle <0$
is sufficiently entangled. In other words, the swapping has the property of
an EW and the following theorem holds true.

\textit{Proposition I: If the expectation value of $\mathbf{S}_{ij}$ on all
separable states is larger or equal to zero, then the inequality
\begin{equation}
\langle \mathbf{S}_{ij}\rangle <0
\end{equation}%
implies that the corresponding quantum state is sufficiently entangled.}

For an SU(2) invariant state of spin-1/2 systems, the condition $%
\left\langle \mathbf{S}_{i,j}\right\rangle <0$ is sufficient and necessary
for entanglement \cite{xgwang}. We would like to emphasize that the above
theorem is not restricted to the spin systems, but also applicable to any
composite systems consisting two identical subsystems, e.g., two $d$-level
systems and two identical infinite-dimensional systems. It is interesting to
notice that Horodecki \textit{et al} had found that any \textit{permutation
of indices} of a density matrix leads to the separability criterion\cite%
{horodecki-2002}. Here our considerations focus on the swapping of the
quantum spin states on two different sites. What is more, our analyses have
shown that \textit{not} all such permutations can be regarded as a
separability criterion.

Swapping operator has appeared in the expression of the concurrence in
spin-1/2 systems, and it can be expected to manifest itself in the
negativity expression of the SU(2)-invariant states for arbitrary quantum
spin-$s$ systems. For an SU(2) invariant state, the density operator can be
written as a linear combination of the projection operators,
\begin{equation}
\rho =\frac{1}{2s+1}\sum_{F=0}^{2s}\frac{\alpha _{F}}{\sqrt{2F+1}}\mathbf{P}%
_{F},\text{ \ }\alpha _{F}=\frac{2s+1}{\sqrt{2F+1}}Tr(\rho \mathbf{P}_{F}),
\end{equation}%
where $F$ is the quantum number of the total angular momentum $(\mathbf{s}%
_{i}\mathbf{+s}_{j})$. After partial transposition with respect to the
second spin, the transposed density matrix still has an SU(2) symmetry, and
can be written as \cite{schliemann}
\begin{equation}
\rho ^{T_{2}}=\frac{1}{2s+1}\sum_{K=0}^{2s}\frac{\alpha _{K}^{\prime }}{%
\sqrt{2K+1}}\mathbf{P}_{K}^{\prime },  \label{eq:pt}
\end{equation}%
where $K$ is the quantum number of another angular momentum composed of $%
\mathbf{s}_{i}$ and $\mathbf{s}_{j}$: $K_{ij}^{x}=s_{i}^{x}-s_{j}^{x}$, $%
K_{ij}^{y}=s_{i}^{y}+s_{j}^{y}$, $K_{ij}^{z}=s_{i}^{z}-s_{j}^{z}$. As shown
by Breuer \cite{breuer}, a relation between the coefficient vectors ${\vec{%
\alpha}}^{\prime }$ and ${\vec{\alpha}}$ can be established%
\begin{eqnarray}
{\vec{\alpha}}^{\prime } &=&\mathbf{\Theta }{\vec{\alpha}},  \notag \\
\mathbf{\Theta }_{FK} &=&\sqrt{(2F+1)(2K+1)}\left(
\begin{array}{ccc}
s & s & F \\
s & s & K%
\end{array}%
\right) ,
\end{eqnarray}%
where $\vec{\alpha}=(\alpha _{0},\alpha _{1},...,\alpha _{2s})^{T}$, $\vec{%
\alpha}^{\prime }=(\alpha _{0}^{\prime },\alpha _{1}^{\prime },...,\alpha
_{2s}^{\prime })^{T}$, and $\mathbf{\Theta }_{FK}$ is given by the Wigner 6-$%
j$ symbol \cite{edmonds}. From Eq.~(\ref{eq:pt}), the negativity of the
corresponding density matrix is then calculated as
\begin{equation}
\mathcal{N}=\frac{1}{2s+1}\sum_{K=0}^{2s-1}\max \left( 0,-\sqrt{2K+1}%
\sum_{F=0}^{2s}\mathbf{\Theta }_{KF}\alpha _{F}\right) ,
\end{equation}%
where the last term in the $K$ summation does not contribute to the
negativity.

For an $s=1/2$ bipartite system, the above negativity gives rise to $%
\mathcal{N}=$max$(0,-\langle \mathbf{s}_{i}\mathbf{\cdot s}_{j}\rangle )$.
However, for the $s=1$ bipartite system, the corresponding negativity is
given by%
\begin{eqnarray}
\mathcal{N} &=&\frac{1}{3}\max (0,-\langle \mathbf{s}_{i}\mathbf{\cdot s}%
_{j}\rangle -\langle \left( \mathbf{s}_{i}\mathbf{\cdot s}_{j}\right)
^{2}\rangle )  \notag \\
&&+\frac{1}{2}\max (0,\langle \left( \mathbf{s}_{i}\mathbf{\cdot s}%
_{j}\right) ^{2}\rangle -2).
\end{eqnarray}
The expectation values of the swapping operators have included in the above
expressions. From the properties of the Wigner 6-$j$ symbol \cite{edmonds},
the first term in the summation over $K$ is given by
\begin{equation}
\Theta _{0F}=(-1)^{2s+F}\frac{\sqrt{2F+1}}{2s+1}.
\end{equation}%
Then, the leading term in the negativity expression can be evaluated as
\begin{align}
& \frac{1}{2s+1}\max \left( 0,(-1)^{2s+1}\sum_{F=0}^{2s}(-1)^{F}\text{Tr}%
(\rho \mathbf{P}_{F})\right)  \notag \\
=& \frac{1}{2s+1}\max \left( 0,-\langle \mathbf{S}\rangle \right) .
\end{align}%
Therefore, being as an EW, the swapping operator has been included in the
expression of negativity as the leading contribution for arbitrary quantum
spin-$s$ systems. Actually, this is also one of the main results of the
present paper.

As an application of the above result, for the following SU(2) invariant
pure state
\begin{equation}
\rho =\frac{1}{4s+1}\mathbf{P}_{2s-1},
\end{equation}%
the expectation value of the swapping operator on this state is found to be $%
-1$, where only the term containing swapping operator survives. Thus, the
negativity for this particular pure state is $1/(2s+1)$, and the
corresponding state is entangled.

\subsection{Generalization of swapping}

A natural generalization of the swapping is the permutation $\mathbf{R}$.
The action of $\mathbf{R}$ on a product state is given by
\begin{equation}
\mathbf{R}|\phi _{1}\rangle \otimes |\phi _{2}\rangle \otimes \ldots \otimes
|\phi _{N}\rangle =|\phi _{i_{1}}\rangle \otimes |\phi _{i_{2}}\rangle
\otimes \ldots \otimes |\phi _{i_{N}}\rangle
\end{equation}%
All $N!$ permutations form a permutation group. We now evaluate $\mathbf{R}$
on a separable state. A $N$-particle density matrix $\rho $ is separable
(non-entangled) if it can be decomposed into
\begin{align}
\rho =& \sum_{k}p_{k}|\phi _{1}^{k}\rangle \langle \phi _{1}^{k}|\otimes
\cdots \otimes |\phi _{i}^{k}\rangle \langle \phi _{i}^{k}|\otimes \cdots
\notag \\
& \otimes |\phi _{j}^{k}\rangle \langle \phi _{j}^{k}|\otimes \cdots \otimes
|\phi _{N}^{k}\rangle \langle \phi _{N}^{k}|,
\end{align}%
where the coefficients $p_{k}$ are positive real numbers satisfying $%
\sum_{k}p_{k}=1$, and $|\phi _{i}^{k}\rangle $ is the state for the $i$-th
particle. For some permutation operators, such as swaps, we can prove that
the corresponding expectation value on a separable state is always large or
equal to zero, we thus conclude that these permutation operators can also be
viewed as EWs. Then, we have following conclusion.

\textit{Proposition II: If the expectation value of permutation $\mathbf{R}$
on all separable states is large or equal to zero, then the inequality
\begin{equation}
\langle \mathbf{R}\rangle <0
\end{equation}%
implies that the corresponding quantum state is sufficiently entangled.}

For $N=2$, the permutation group contains a swap and an identity. For $N=3$,
the permutation group contains 6 elements, and three different swappings,
namely, $\mathbf{S}_{12}$, $\mathbf{S}_{13}$, and $\mathbf{S}_{23}$ are EWs.
For $N=4$, there are 24 elements, and except swappings, there are other
permutations can be viewed as EWs, e.g., $\mathbf{S}_{12}\mathbf{S}_{34}$, $%
\mathbf{S}_{13}\mathbf{S}_{24}$, and $\mathbf{S}_{14}\mathbf{S}_{23}$. Among
them, the operator $\mathbf{S}_{14}\mathbf{S}_{23}$ can be viewed as an
mirror reflection. Furthermore, any superpositions of the EWs $%
\sum_{k=1}^{M}c_{k}\mathbf{P}_{k}$ with $c_{k}$ being positive can also be
viewed as new EWs.

\subsection{Singlet projector as an EW}

For the SU(2) invariant states, the negativity for the spin-1 bipartite
systems has been obtained as
\begin{equation}
\mathcal{N}=\frac{1}{2}\max \left( 0,3\langle \mathbf{P}_{i,j}\rangle
-1\right) +\frac{1}{3}\max \left( 0,-\langle \mathbf{S}_{ij}\rangle \right) .
\end{equation}%
We have observed that the inequality $\langle \mathbf{P}_{i,j}\rangle >1/3$
also implies that the corresponding state is entangled. As we have shown in
previous section, the spin swapping operator and singlet projector are
independent though the partial transposition is related to them. For the
SU(2)-invariant state, there can be two \textit{different} sufficient
entanglement conditions for the spin-1 bipartite systems: one is $%
\left\langle \mathbf{S}_{i,j}\right\rangle <0$ and another is $\left\langle
\mathbf{P}_{i,j}\right\rangle >1/3$. In fact, we can prove a more general
theorem for arbitrary quantum spin-$s$ systems.

\textit{Proposition III: If the expectation value of the singlet projector
satisfies
\begin{equation}
\left\langle \mathbf{P}_{i,j}\right\rangle >\frac{1}{2s+1},
\end{equation}%
the corresponding many-body quantum spin state is sufficiently entangled.}

Proof: A singlet state is given by
\begin{equation}
|\Psi _{s}\rangle =\frac{1}{\sqrt{2s+1}}\sum_{m=-s}^{s}(-1)^{s-m}|s,m\rangle
\otimes |s,-m\rangle ,
\end{equation}%
and the singlet projector can be expressed as $P_{0}=|\Psi _{s}\rangle
\langle \Psi _{s}|$. A product state can always be written as
\begin{equation}
|\Phi \rangle =|\Phi _{1}\rangle \otimes |\Phi _{2}\rangle
=\sum_{m,m^{\prime }}a_{m}b_{m^{\prime }}|s,m\rangle \otimes |s,m^{\prime
}\rangle ,
\end{equation}%
where $\sum_{m}\left\vert a_{m}\right\vert ^{2}=\sum_{m}\left\vert
b_{m}\right\vert ^{2}=1$. Then the expectation value $\langle P_{0}\rangle $
with respect to this product state is found to be
\begin{equation}
\langle \mathbf{P}_{ij}\rangle =\frac{1}{2s+1}\left\vert
\sum_{m=-s}^{s}(-1)^{s-m}a_{m}b_{-m}\right\vert ^{2}\leq \frac{1}{2s+1},
\label{eq:p}
\end{equation}%
where the inequality follows from the Schwatz inequality and the
normalization conditions. We may easily extend the above inequality to the
case of any separable state. For an arbitrary separable state $\rho
_{sep}=\sum_{k}p_{k}\rho _{k}$ with $\rho _{k}$ being the product state. The
expectation value of $\mathbf{P}_{ij}$ satisfies the inequality

\begin{equation}
\langle \mathbf{P}_{ij}\rangle =\text{Tr}(\mathbf{P}_{ij}\rho )=\sum_{k}p_{k}%
{\text{T}r}(\mathbf{P}_{ij}\rho _{k})\leq \frac{1}{2s+1},
\end{equation}%
where we have used Eq.~(\ref{eq:p}). Therefore, the theorem has been proved,
and at the same time the operator $\left( \mathbf{P}_{ij}-\frac{1}{2s+1}%
\right) $ is another class of EW.

\subsection{Relations with other EWs}

The quantum spin Hamiltonians have already been used as EWs to detect
entanglement~\cite{EW_Dowling,EW_Toth}. Here, we would like to study the
relations among them. Let us consider the following Hamiltonian
\begin{equation}
H=J\sum_{\langle i,j\rangle }\mathbf{S}_{i,j},
\end{equation}%
which is a sum of all different swaps on the nearest neighbor sites. We know
that every expectation value of each swap on a separable state is large or
equal to zero. Then, the expectation value of the Hamiltonian on a separable
state satisfies $\langle H\rangle \geq 0$. Therefore, the Hamiltonian is
regarded as an EW too. For any eigenstate, if the eigenenergy is less than
zero, the many-body state must be entangled. We see that a new EW was
constructed by superpositions of swaps. In fact, any superposition of swaps
with positive coefficients are EWs as well. Furthermore, it is more
interesting to consider some other models consisting of the swapping
operators with supersymmetries \cite{korepin}.

Similarly, we consider the following Hamiltonian in terms of the singlet
projections
\begin{equation}
H=J\sum_{\langle i,j\rangle }\left( \mathbf{P}_{i,j}-\frac{1}{2s+1}\right)
\end{equation}%
From the proposition II, we can easily prove that $\langle H\rangle \geq 0$
for a separable state, indicating that the Hamiltonian can be viewed as an
EW. Any superposition of operators ${\mathbf{\tilde{P}}_{i,j}=}\left(
\mathbf{P}_{i,j}-\frac{1}{2s+1}\right) $ with positive coefficients are EWs
as well. Moreover, the

\section{Summary}

We have derived the general form of the spin swapping operator for the
quantum Heisenberg spin-$s$ systems, and proved that under the partial
transposition the general spin swapping operator is equal to the singlet
projection operator. For SU(2) invariant bipartite spin-$s$ systems, we also
found that the expectation value of the swapping operator is the leading
contribution to the negativity. Generalized to the many-body particle
states, the expectation values of the swapping and permutation operators can
be used as an entanglement witness, which, moreover, in some cases can be
used for formulation of necessary and sufficient condition of entanglement.
This is a quite important and new mathematical fact, which could be used for
the formulation of observable conditions of entanglement in near future.

The authors acknowledged that this research work was finalized when both of
us visited the Center for Theoretical and Computational Physics of the
University of Hong Kong. G. M. Zhang is supported by NSF-China (Grant No.
10125418 and 10474051). X. G. Wang is supported by NSF-China under grant no.
10405019, Specialized Research Fund for the Doctoral Program of Higher
Education (SRFDP) under grant No.20050335087, and the project-sponsored by
SRF for ROCS and SEM.

\end{document}